\documentclass[twocolumn,superscriptaddress,showpacs,preprintnumbers,amsmath,amssymb,oneside,prl]{revtex4}
\usepackage{graphicx}
\usepackage{dcolumn}
\usepackage{bm}

\begin{document}
\title{Experimental Quantum Teleportation and Multi-Photon Entanglement via Interfering Narrowband Photon Sources}
\author{Jian Yang}
\affiliation{Hefei National Laboratory for Physical Sciences at
Microscale and Department of Modern Physics, University of Science
and Technology of China, Hefei, Anhui 230026, China}

\author{Xiao-Hui Bao}
\affiliation{Physikalisches Institut der Universitaet Heidelberg,
Philosophenweg 12, Heidelberg 69120, Germany}
\affiliation{Hefei National Laboratory for Physical Sciences at
Microscale and Department of Modern Physics, University of Science
and Technology of China, Hefei, Anhui 230026, China}

\author{Han Zhang}
\affiliation{Hefei National Laboratory for Physical Sciences at
Microscale and Department of Modern Physics, University of Science
and Technology of China, Hefei, Anhui 230026, China}

\author{Shuai Chen}
\affiliation{Hefei National Laboratory for Physical Sciences at
Microscale and Department of Modern Physics, University of Science
and Technology of China, Hefei, Anhui 230026, China}
\affiliation{Physikalisches Institut der Universitaet Heidelberg,
Philosophenweg 12, Heidelberg 69120, Germany}

\author{Cheng-Zhi Peng}
\affiliation{Hefei National Laboratory for Physical Sciences at
Microscale and Department of Modern Physics, University of Science
and Technology of China, Hefei, Anhui 230026, China}

\author{Zeng-Bing Chen}
\affiliation{Hefei National Laboratory for Physical Sciences at
Microscale and Department of Modern Physics, University of Science
and Technology of China, Hefei, Anhui 230026, China}

\author{Jian-Wei Pan}
\affiliation{Hefei National Laboratory for Physical Sciences at
Microscale and Department of Modern Physics, University of Science
and Technology of China, Hefei, Anhui 230026, China}
\affiliation{Physikalisches Institut der Universitaet Heidelberg,
Philosophenweg 12, Heidelberg 69120, Germany}

\date{\today}

\begin{abstract}
In this letter, we report a realization of synchronization-free
quantum teleportation and narrowband three-photon entanglement
through interfering narrowband photon sources. Since both the single-photon and the entangled photon pair utilized are completely autonomous, it removes the requirement of high demanding synchronization technique in long-distance quantum communication with pulsed spontaneous parametric down-conversion sources. The frequency linewidth of the three-photon entanglement realized is on the order of several MHz, which matches the requirement of atomic ensemble based quantum memories. Such a narrowband multi-photon source will have applications in some advanced quantum communication protocols and linear optical quantum computation.
\end{abstract}

\pacs{03.67.Bg, 42.65.Lm}

\maketitle

Quantum teleportation \cite{teleportation_Bennett} is a process to
transfer a quantum state of a photon without transferring the state
carrier itself, which plays a central role in quantum communication
\cite{GISINrmp}. It necessitates the interference of a single-photon
and an entangled photon pair. Since spontaneous parametric
down-conversion (SPDC) is the main method to generate entangled
photons \cite{Kwiat}, typically with a frequency linewidth of
several THz. To interfere independent sources, the resolution time
of the photon detectors has to be much smaller than the coherence time ($<1\,\rm{ps}$)\cite{swapping}, which is
still not available until today. This problem was later solved by
utilizing a femtosecond pumping laser and frequency filtering
\cite{Ind_pulse_Zukowski,teleportation_first}. Since the development of this technique,
numerous important advances have been achieved
\cite{swapping_exp,pan2000,pan2003pur,Walther2004,Walther2005}. But
in this pulsed regime, interference of independent sources requires
a synchronization precision of several hundred fs for the pumping
lasers. Even though there are some experimental investigations
\cite{synch_Yang, synch_Zeilinger} with lasers within a single lab,
when one wants to build entanglement over several hundred kilometers, it will become rather challenging.

While in the continuous-wave regime, with the development of the
quasi-phase matching technique, it is now possible to narrow the
frequency bandwidth for SPDC sources to several tens GHz
\cite{Konig2005}, lowering down the requirement for photon
detectors. In \cite{narrow_Gisin} Halder \textsl{et al.} has
demonstrated the feasibility to interfere separate sources through
time measurement. In their experiment, Bragg gratings were used to
filter out narrow-band photons from a SPDC source, increasing the
coherence time to several hundred ps. In order to interfere such
entangled sources, a high-demanding superconducting detector with
ultra-low time jitter was utilized, which is only available for few
groups. Recently, we have reported a narrow-band entangled photon
source with a $\sim$MHz linewidth through cavity-enhanced SPDC
\cite{narrowband}. Such a narrow-band source will enable the
possibility to interfere separate sources with the widely used
commercial sub-ns photon detectors. Also the tolerance of length
fluctuations for the quantum communication link will improve from
several centimeters in \cite{narrow_Gisin} to several meters, which
means we can realize quantum teleportation for longer distance,
larger time scale, and worse weather condition.

Interference of independent sources is also the main method to
generate multi-photon entanglement
\cite{5-photon_Z.Zhao,lu2007,pan2008}, which is the main resource
for linear optical quantum computation (LOQC)
\cite{klm2001}. To efficiently build
large entangled states for LOQC, it is required to store the intermediate
multi-photon entangled states with a quantum memory \cite{cluster_Rudolph,duan}. But previously
due to the usage of SPDC sources, the frequency linewidth of these
multi-photon entanglement lies on the order of several THz. While
the frequency linewidth required by an atomic ensemble based quantum
memory \cite{DLCZ,QM_Kimble,QM_Kuzmich,QM_Pan} is on the order of
several MHz. This frequency mismatch greatly limits the applications
of the broadband multi-photon entangled sources. Therefore, creating a narrowband multi-photon entanglement with linewidth of several MHz
becomes an urgent task.

\begin{figure*}[hbtp]
\includegraphics[width=13cm]{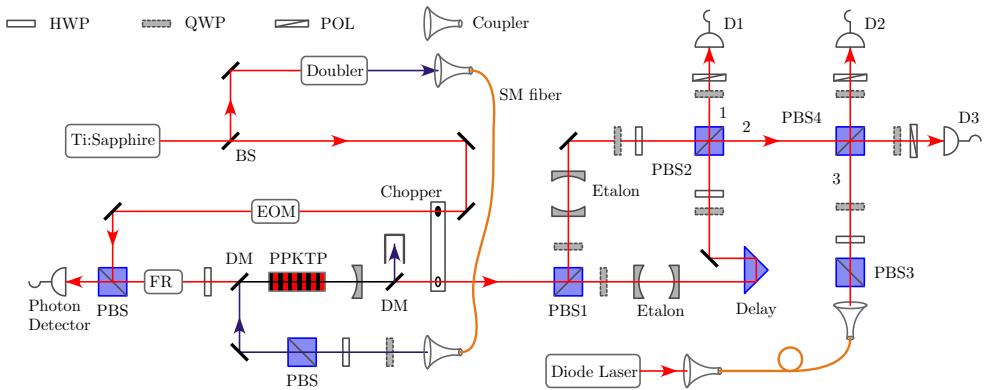}
\caption{Experimental setup.  A continuous-wave Ti:Sapphire laser locked to one of the $\rm{^{87}Rb}$ $\rm{D_2}$ hyperfine transitions is divided into two portions
by a beam-splitters (BS). The main portion is up-converted to an ultraviolet (UV) beam (390nm) through an external-cavity doubler.
Then the UV pump beam (about 9 mw) is coupled into the double-resonant cavity which is composed of a PPKTP (from Raicol) and a concave mirror. The PPKTP is type-II configured that one UV photon gives rise to a $H$ polarized photon and a $V$ polarized photon. After separated on polarized beam-splitter 1 (PBS1), an etalon on each arm is used to filter out single longitudinal mode.
The quarter-wave plate (QWP) before each etalon is used to form an optical isolator with PBS1 to eliminate the reflection from the etalon. After rotated to $|+\rangle$ polarization with a half-wave plate (HWP) respectively, the two photons get interfered on PBS2
.The narrow-band entangled photon pair is generated in the case that one photon appears on each output port. The second portion is used to lock the double-resonant cavity actively. A Faraday rotator (FR) and a PBS are utilized to extract the reflected locking beam in order to generate the error signal for the locking. The optical chopper is utilized to switch between the locking and the detecting process.
The beam from another completely independent diode laser (locked to the same atomic transition line as the Ti:Sapphire laser) is attenuated to an intensity of about $8.0\times 10^5 \, \mathrm{s}^{-1}$ as the single-photon source to be teleported. A partial Bell state measurement (BSM) of photon 2 and photon 3 is realized with PBS4 and the following polarization analyzers.}\label{setup}
\end{figure*}

In this Letter, we experimentally investigate the interference of a
single-photon and an entangled photon pair, both of which are
continuous-wave and narrowband ($\sim$MHz). Through this
interference, first we realize a synchronization-free quantum
teleportation. Since both for the single-photon and the entangled
photon pair utilized are completely autonomous, it removes the requirement of high demanding synchronization technique for the case of pulsed SPDC sources, enabling the possibility to teleport a photonic state between distant locations. Secondly the same setup enables us to generate a narrowband three-photon entangled state, with a linewidth of several MHz, which matches the requirement of atomic ensemble based quantum memories. Such a narrowband multi-photon source will have applications in some advanced quantum communication protocols \cite{secret_sharing, 5-photon_Z.Zhao} and LOQC.

The schematic diagram of our experimental setup is shown in Fig. \ref{setup}. The entangled photon pairs are generated through cavity-enhanced SPDC. Measured linewidth for this
source is 9.6 MHz. Single-mode output is realized by setting a
cavity length difference between different polarizations, active
cavity stabilization and the use of temperature controlled etalons.
Detailed description of this narrow-band entangled source could be
found in a former paper of us \cite{narrowband}. The quantum state
of these two photons can be expressed as:
\begin{eqnarray}
|\Phi^{-}\rangle_{12}&=&\frac{1}{\sqrt{2}}(|H\rangle_1|H\rangle_2 -
|V\rangle_1|V\rangle_2),
\end{eqnarray}
which is one of the four Bell sates, and $H$ represents horizontal polarization, $V$ represents vertical polarization. For a 9 mw input UV power, it is observed that a two-fold coincidence rate of about $200\,
\mathrm{s}^{-1}$ at a coincidence time-window of 16 ns.

The single-photon to be teleported (photon 3 in Fig. \ref{setup}) is
generated by attenuating another completely independent diode laser to an intensity of about $8.0\times 10^5 \,
\mathrm{s}^{-1}$. The state to be teleported is prepared with a HWP
or a QWP. A partial Bell state
measurement (BSM) is realized by sending photon 2 and photon 3
through the PBS4 and the following polarization analyzers. When a
coincidence of $|+\rangle_2|+\rangle_3$
($|\pm\rangle=\frac{1}{\sqrt{2}}(|H\rangle\pm|V\rangle)$) clicks
between detector D2 and D3, the two photons are projected into the
state of $|\Phi^{+}\rangle_{23}=1/\sqrt{2}(|H\rangle_2|H\rangle_3 +
|V\rangle_2|V\rangle_3)$. Then after a local operation of $\sigma
_z$ on photon 1, the teleportation from photon 3 to photon 1 is
finished. In order to get a high-visibility interference on PBS4
between the single-photon and the entangled pair, the coincidence
time window between photon 2 and photon 3 should be much smaller
than the correlation time between photon 1 and photon 2 (20ns)
\cite{Ind_pulse_Zukowski}, in our case we choose it to be 3 ns.

\begin{figure}[hbtp]
\includegraphics[width=1\columnwidth]{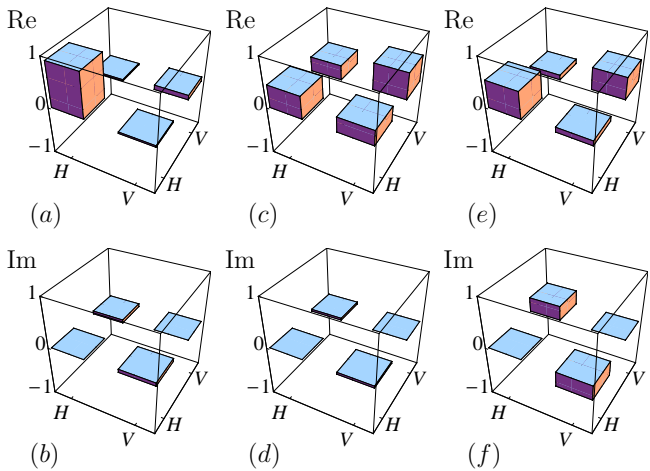}
\caption{Tomography results for the teleportation of $|H\rangle$, $|+\rangle$ and $|\textrm{L}\rangle$. (a) and (b) are
the real and imaginary parts for the teleported state of $|H\rangle$ respectively. (c) and (d) are the real and imaginary parts for the teleported state of $|+\rangle$ respectively. (e) and (f) are the real and imaginary parts for the teleported state of $|L\rangle$ respectively.
}\label{tomo}
\end{figure}

\begin{table}[!hbp]
\centering \caption{Fidelities for the teleportation experiment. All the fidelities of the teleportion are well above
the classical limit of 2/3.} \label{table1}
\begin{tabular*}{1\columnwidth}{@{\extracolsep{\fill}}lccc}
\hline \hline Polarization & $|H\rangle$ & $|+\rangle$
& $|L\rangle$\\
\hline
Fidelity & 91.0\% & 79.8\% & 79.0\%\\
\hline
\end{tabular*}
\end{table}

For the states to be teleported of photon 3, we choose three states, namely, $|H\rangle$, $|+\rangle$, and left-handed ($|\textrm{L}\rangle$) circular polarization states $\frac{1}{\sqrt{2}}(|H\rangle - i |V\rangle)$. In order to evaluate the performance for the teleportation
process, we make a quantum tomography \cite{tomography} for all the
teleported states, with results shown in Fig.
\ref{tomo} and fidelities shown in Table.
\ref{table1}. It shows that the fidelities of the six states are well above the classical limit of 2/3 \cite {classical_limit}. Thus the success of quantum teleportation is proved.

A great advantage to use continuous-wave sources for quantum
teleportation and entanglement connection is that the two sources
can be completely autonomous. In our experiment, for a 3 ns
coincident time window, which is much smaller than the coherent time
of the input photons, a perfect overlap between photon 2 and photon
3 at PBS4 is always guaranteed. It removes the high-demanding synchronization technique and provides a much easier way to generate entanglement by using completely independent sources over a large distance.

\begin{figure}[hbtp]
\includegraphics[width=7cm]{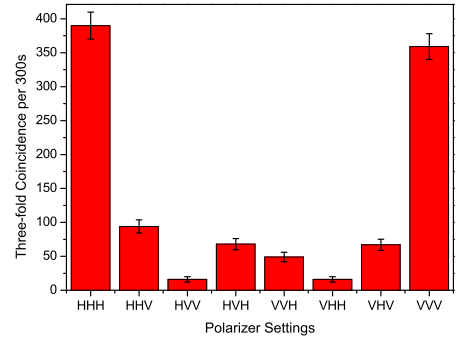}
\caption{Measured result for the three-photon entangled state in $H/V$ basis. It shows that the signal-to-noise
ratio between the desired three-fold components
to any of the 6 other non-desired ones, is about 7.3:1, which
confirms that $HHH$ and $VVV$ are the main components of the
three-photon state. Error bars represent the statistical errors.}\label{component}
\end{figure}

With similar setup\cite{mynote}, it is now possible to generate the first narrow-band three-photon entanglement. By preparing photon 3 in the state of $|+\rangle$ and photon 1 and 2 in the entangled state of $|\Phi^-\rangle$, the three-fold coincidence among the detector D1,
D2 and D3 will lead to a three-photon GHZ state \cite{GHZ}:
\begin{equation}\label{three_photon}
|\Phi\rangle_{123}=\frac{1}{\sqrt{2}}\left(|H\rangle_1|H\rangle_2|H\rangle_3-
|V\rangle_1|V\rangle_2|V\rangle_3 \right).
\end{equation}
To experimentally verify that the desired state of Eq.
\ref{three_photon} has been successfully generated, we first
characterize the components of the three-photon state corresponding to such a three-fold coincidence. This was done by measuring each
photon in the $H/V$ basis. The result is shown in Fig.
\ref{component}. The signal-to-noise ratio, which is defined as the
ratio of any of the desired three-fold components ($HHH$ and $VVV$)
to any of the 6 other non-desired ones, is about $7.3:1$.

To obtain a further characterization of the entanglement, we  make a
measurement of the Mermin inequality \cite{Ineq_Mermin}. In our
case, the value of $A$ in the inequality is defined as:
\begin{equation}\label{Mermin}
A=\sigma_y^1\sigma_y^2\sigma_x^3+\sigma_y^1\sigma_x^2\sigma_y^3+\sigma_x^1\sigma_y^2\sigma_y^3-\sigma_x^1\sigma_x^2\sigma_x^3
\end{equation}
where $\sigma^j_i$ corresponds to the $i$ th Pauli matrix on
particle $j$. Violation of the inequality , that is $|\langle A
\rangle|>2$, proves the non-local property of the three photon
state. The measured value of the observables are shown in Table.
\ref{table2}. With simple calculation, it is obtained that $|\langle
A \rangle|=2.59\pm0.05$, which violates the inequality by 12
standard deviations. Combining the results of components and Mermin
obserbales measurements, we can obtain the fidelity between the
state generated and the ideal three-photon GHZ state:
\begin{eqnarray}
F(\rho)&=&
\frac{1}{2}(_{123}\langle HHH |\rho| HHH \rangle_{123}
\nonumber\\
&&+_{123}\langle VVV |\rho| VVV \rangle_{123})+\frac{1}{8}|\langle
A\rangle|
\end{eqnarray}
\begin{table}
\centering \caption{Measured observables of the Mermin inequality
(Eq. \ref{Mermin}) for the three-photon entangled state.}
\label{table2}
\begin{tabular*}{1\columnwidth}{@{\extracolsep{\fill}}lcccccc}
\hline \hline Observable & $\sigma_y^1\sigma_y^2\sigma_x^3$ &
$\sigma_y^1\sigma_x^2\sigma_y^3$ & $\sigma_x^1\sigma_y^2\sigma_y^3$
& $\sigma_x^1\sigma_x^2\sigma_x^3$\\
\hline
Value & 0.64 & 0.63 & 0.67 & -0.66\\
\hline
Deviation & 0.02 & 0.02 & 0.02 & 0.02\\
\hline
\end{tabular*}
\end{table}
Our result is $F(\rho)=0.68\pm0.01$, which is well above the
boundary of 1/2, and thus a proof of true three-photon entanglement
\cite{fidelity_threshold}. As the linewidth of entangled
three-photon is of several MHz, it may have broad application in
future LOQC together with atomic quantum memory, especially for the generation of large cluster states \cite{cluster_Rudolph} that are storable.

In summary, a realization of synchronization-free quantum
teleportation and narrowband three-photon entanglement through
interfering continuous-wave narrowband sources is reported. Since
both for the single-photon and the entangled photon pair utilized
are completely autonomous, it removes the requirement of high
demanding synchronization technique for the case of pulsed SPDC
sources, enabling the possibility to teleport a photonic state
between distant locations. The frequency linewidth of the narrowband
three-photon entanglement realized is on the order of several MHz,
which matches the requirement of atomic ensemble based quantum
memories. Such a narrowband multi-photon source will have
applications in some advanced quantum communication protocols and
LOQC.

We thank Yong Qian and Xian-Min Jin for help and discussion. This
work is supported by the NNSF of China, the CAS, and the National
Fundamental Research Program (under Grant No. 2006CB921900) and the
ERC grant.

\bibliography{myref}

\begin{thebibliography}{33}
\bibitem[{myn()}]{mynote}
\bibinfo{note}{For part it is not necessary to make the entangled photon pair
  and the single photon independent from each other, so photon 3 is got by
  attenuating the beam from the Ti:Sapphire laser for conveniency.}
\end{thebibliography}

\end{document}